\documentstyle[twocolumn,aps,psfig,epsf]{revtex}
\begin{document}
\draft
\twocolumn[\hsize\textwidth\columnwidth\hsize\csname @twocolumnfalse\endcsname
%

\title{Auxiliary--field Monte Carlo for Quantum Spin and Boson Systems}

\author{M.~Ulmke$^1$ and R.~T.~Scalettar$^2$}

\address{$^1$Theoretische Physik III, EKM,
Institut f\"ur Physik, Universit\"at Augsburg, 
D 86135 Augsburg, Germany\\
$^2$Department of Physics, University of California, Davis, CA 95616, USA}

\date{\today}
\maketitle

\begin{abstract} 

We describe a new algorithm for the numerical simulation
of quantum spin and boson systems. The method is based on the 
Trotter decomposition in imaginary time and a decoupling
by auxiliary Ising spins.
It can be applied, in principle, to arbitrary (random)
spin systems, however in general it suffers from the
``minus-sign problem''. This problem is absent in the
case of the Ising model in a transverse field in arbitrary
dimensions and geometries.
We show test results for the spin--1/2 XY model,
the one--dimensional transverse Ising model with disorder,
and the phase transition 
induced by a transverse field in the two-dimensional 
Ising model.  

\end{abstract}

\vskip2pc]

\section{Introduction}

Quantum Monte Carlo (QMC) methods have contributed to much of our recent
knowledge of the properties of interacting quantum mechanical
spin systems, and closely related hard--core boson models.
Green Function Monte Carlo (GFMC)
is one powerful class of 
approaches which project out the lowest energy many--body eigenfunction.
World--line (WLQMC) algorithms constitute another generic class
of QMC approaches, and allow the evaluation of finite temperature
properties.
Recently, very substantial improvements to WLQMC, the loop\cite{LOOP}
and continuous time\cite{CONTTIME} techniques,
have been developed.

Both GFMC and WLQMC most commonly use a
basis labeled by the boson occupation number or position,
or, analogously, the $z$ component of spin,
in space and imaginary time.
The key feature of the approaches is that eigenvalues
of the original operators in the Hamiltonian describe the
Monte Carlo configurations.
In contrast, the preferred techniques for QMC 
simulations of interacting fermions\cite{DET},
involve the introduction of an auxiliary field.
The original fermion operators are integrated out, and
the simulation takes place in the space of this
abstract auxiliary field.

In this paper we will introduce a new auxiliary
field QMC method for interacting quantum mechanical
spins and boson systems.
Why is such an algorithm interesting?
WLQMC and GFMC approaches have very significant
weaknesses, including extremely long correlation times 
and restrictions on the observables which can be measured.
While loop algorithms\cite{LOOP} have addressed this issue, 
their efficiency remains
problematic in several important cases, for example when interactions
are longer range, or disorder is present.
Therefore, continued algorithm development is desirable.

The organization of this paper is as follows:  We will first
introduce the Ising model in a transverse field,
and briefly review the key issues in its properties.  
We will then describe how an auxiliary field
algorithm can be constructed for this model. 
Although related conceptually to fermion QMC, 
it differs considerably from analogous fermion techniques
in that the resulting traces are over independent single site
problems, avoiding the necessity to evaluate the determinants
of large matrices in the fermion case.
We then give results
of our simulations, including a comparison of
the approach with existing techniques.
We conclude by describing another interacting
spin/boson model, the boson--Hubbard model.
However, we show that the sign problem is a serious limitation 
to auxiliary field approaches in this case.

\section{Transverse Field Ising Model}

Two quantum spin/hard--core boson problems of 
considerable recent issue are 
the Ising model in a transverse field\cite{YOUNG},
and the boson--Hubbard model\cite{FISHER}.
The former allows one
to study in a simple setting many of the key qualitative
issues in quantum phase transitions in disordered
systems, including the nature of the distribution
of correlation functions and the shifts in the values
of critical exponents from the clean limit.
The latter offers a description of the superconductor--insulator
phase transition when preformed pairs exist above the
transition, and in the hard--core limit is also formally identical to the
quantum mechanical spin-1/2 XXZ Hamiltonian.
In this section we will describe the Ising model in a transverse
field, which appears to be the more promising application of
the auxiliary field approach.  

\subsection{Hamiltonian and Algorithm}

The Transverse Ising model \cite{YOUNG,Chakrabarti96} is given by
\begin{equation} 
H = \sum_{\langle ij \rangle} J_{ij} S_i^z S_j^z - \sum_i B_i S_i^x.
\end{equation} 
Here $S_i^\alpha$, $\alpha\in\{x,y,z\}$, are the Pauli matrices
obeying the commutation relations: 
$[S_i^\alpha,S_j^\beta]=\delta_{ij} (\epsilon_{\alpha\beta\gamma}
S_i^\gamma+\delta_{\alpha\beta} ).$
The sum runs over pairs of nearest neighbors, $\langle ij \rangle$.
The partition function is given by  
${\cal Z}= {\rm Tr}\; e^{-\beta H}$.

We employ the usual Trotter break-up of non-commuting operators
in the Hamiltonian:

\begin{eqnarray}
&& e^{-\beta H}  =  \left( e^{-\Delta\tau H}\right)^L  \label{Trotter} \\
&& \approx  \left[ \prod_{i=1}^{N} \hspace{-0.2em}
\left( \prod_{j\in {\rm NN}(i)}  \hspace{-0.4em}
e^{-\Delta\tau J_{ij}S_i^z S_j^z}\right)
 e^{\Delta\tau B_i  S_i^x} + {\cal O}(\Delta\tau^2) \right] ^L,
\nonumber 
\end{eqnarray}
with $\Delta\tau L=\beta$. The inner product runs over the $Z$ nearest
neighbors of $i$.

In order to decouple the interaction terms we recall that any product
of two commuting operators can be written as a sum over squares:
\begin{equation}
AB=\frac{1}{4} \left[ (A+B)^2 - (A-B)^2 \right],
\end{equation}
and a squared operator can be decoupled by the introduction of
a Gaussian integration over a classical auxiliary field:
\begin{equation}
e^{(A+B)^2/4} = \pi^{-1/2} \int dx \; e^{-x^2+(A+B) x}.
\end{equation}
In the present case, one squared operator is sufficient
\begin{equation}
S_i^z S_j^z = \alpha \left[ 2 (P_{ij}^\alpha)^2-1\right], \;  {\rm with}\; 
P_{ij}^\alpha= (S_i^z +\alpha S_j^z)/2
\end{equation}
and $\alpha=\pm 1$. 

Employing that $(P_{ij}^\alpha)^2$ is a 
projection operator, 
i.e.~$(P_{ij}^\alpha)^{2k}=(P_{ij}^\alpha)^2$, for $k=1,2,\cdots$,
one immediately confirms by Taylor expansion that:

\begin{eqnarray}
e^{-2\alpha\Delta\tau J_{ij} (P_{ij}^\alpha)^2 } & = & 
\cosh \left(  2\lambda_{ij} P_{ij}^\alpha \right) \nonumber \\
&=&\frac{1}{2}\sum_{\sigma=\pm1}  e^{\sigma\lambda_{ij} (S_i^z+\alpha S_j^z)}
\end{eqnarray}
with $\cosh(2\lambda_{ij})=\exp({-2\alpha\Delta\tau J_{ij}})$.
In order to get real variables one chooses $\alpha=+1$ ($-1$)
for $J_{ij}<0$ ($>0$).
Thus a two-valued rather than Gaussian decoupling of the interaction
is possible, introducing Ising-type auxiliary spins $\sigma_{ij}=\pm 1$:

\begin{equation}
 e^{-\Delta\tau J_{ij}S_i^z S_j^z} = \frac{1}{2} \; e^{-\Delta\tau J_{ij}}  
\sum_{\sigma_{ij}=\pm 1} e^{\lambda_{ij}
\sigma_{ij} (S_i^z - S_j^z)}
\label{HStrafo}
\end{equation}
for $J_{ij}>0$.
This ``discrete Hubbard--Stratonovich transformation''
was first introduced by Hirsch in the fermion case\cite{HIRSCH}.

The same decoupling holds for any component of the Pauli matrix,
hence the extension to XY or (anisotropic) Heisenberg models or,
equivalently, hard-core bosonic models, as will be discussed.
 
The decoupling has to be done for each of the $L$ factors in (\ref{Trotter})
giving the auxiliary spins $\sigma_{ij}(l)$ an additional index
$l=1, \cdots, L$.
As a result, we have a system of non-interacting Ising spins
in a transverse field $B_i$ coupled to an auxiliary 
longitudinal two-valued field, described by $\sigma_{ij}(l)$, 
which fluctuates in space and ``imaginary time''. For a given
configuration, $\{\sigma_{ij}(l)\}$, the original spins are trivially 
described by a direct product of $2\times 2$ matrices.

Note that $\sigma_{ij}(l)$ is a \em bond, \em not site variable.
$i$ and $j$ denote pairs of sites connected by a $J_{ij}\ne 0$.
Hence, there are $NZL/2$ auxiliary spins where $Z$ is the coordination
number of the lattice. Even in the classical limit ($B_i\equiv 0$, L=1)
the auxiliary spins are \em not \em dual to the original ones.
An exception is the one-dimensional classical case where 
original and auxiliary spins are equivalent.

As mentioned above, for a given configuration of auxiliary spins
the original spins are independent, the Hilbert space factorizes,
and the partition function can be written as:

\begin{eqnarray}
{\cal Z} & = & \sum_{\{\sigma_{ij}(l)\}} \prod_{i=1}^N 
{\rm Tr}_i  \underbrace{  \prod_{l=1}^L \left(
\prod_{j\in {\rm NN}(i)} e^{\lambda_{ij}\sigma_{ij}(l) S_i^z}
\; e^{\Delta\tau B_i S_i^x} 
\right)}_{2\times 2\; {\rm matrix \; product}} \label{partitionfct}
\\
&&\nonumber\\
&= & \sum_{\{\sigma_{ij}(l)\}}  
\underbrace{w(\{\sigma_{ij}(l)\})}_{\qquad 
\rm positive\; definite\; weight\; function} 
\label{weight}
\end{eqnarray}
That is, 
${\cal Z}$ is now a sum over the $NZL/2$ auxiliary Ising spins with a weight
function proportional to the product of traces of $2\times 2$ matrices,
one for each lattice site.

In the case of the Ising model one can choose all local transverse 
magnetic field values $B_i$ to be positive or change their sign by
a local spin rotation, respectively.
Thus there are only positive matrix elements involved and 
$w(\{\sigma_{ij}(l)\})$ can serve as a positive definite weight function.
This is, however, different, e.g.~in the XY model where severe sign
problems occur even for small systems (see Sec.~III).

To simplify the notation for the following,
we denote every factor in the matrix product (\ref{partitionfct}) by 
\begin{equation}
{\cal B}_i(l) \equiv  
\left( \prod_{j\in {\rm NN}(i)} e^{\lambda_{ij}\sigma_{ij}(l) S_i^z}
\right) e^{\Delta\tau B_i S_i^x} \label{Bmatrix}
\end{equation}
and define the product over $l$ and its cyclic permutations as 
\begin{equation}
{\cal A}_i(l) \equiv  {\cal B}_i(l) \;
{\cal B}_i(l+1) \cdots {\cal B}_i(L) \; {\cal B}_i(1) \cdots {\cal B}_i(l-1) 
\label{Amatrix}
\end{equation}
In fact,
the values of the traces do not change under cyclic permutation
in the matrix product, i.e.~they do not depend on the index $l$ 
in $ {\cal A}_i(l)$,  and we can rewrite ${\cal Z}$ as:
\begin{equation}
{\cal Z}  =  \sum_{\{\sigma_{ij}(l)\}} \prod_{i=1}^N 
{\rm Tr}_i \; {\cal A}_i(1) 
\end{equation}

The resulting Monte Carlo algorithm is similar to the auxiliary--field methods 
for lattice fermions \cite{DET},
with the key difference that one has the product of traces
of $NZL/2$ matrices of dimension 2 to evaluate, instead of the determinant of
a single matrix of dimension the spatial lattice size $N$. 
Thus the algorithm scales linearly in $N$.
It goes as follows:

\begin{enumerate}
\item One starts at ``time slice'' $l=1$, initializes the auxiliary field
$\sigma_{ij}(l)$, and calculates the $2\times 2$ matrix products 
${\cal A}_i(l)$ for each lattice site. 
The possible matrix elements $(\exp(\pm \lambda_{ij} S_i^z),\; 
\exp(\Delta\tau B_i S_i^x))$ 
needed to determine ${\cal B}_i(l)$ in (\ref{Bmatrix}) are calculated
once for all at the beginning.
\item At a fixed time slice $l$ try $NZ/2$ single-spin flips,
$\sigma_{ij}(l)\to \sigma'_{ij}(l)=-\sigma_{ij}(l)$. 
Such a flip involves the two 
lattice sites $i$ and $j$, and is accepted according to the 
probability ratio
\begin{equation}
p = \frac{{\rm Tr} \; {\cal A}'_i(l) \; {\rm Tr} \; {\cal A}'_j(l)}
{{\rm Tr} \; {\cal A}_i(l) \; {\rm Tr} \; {\cal A}_j(l)},
\end{equation}
with the new matrices
\begin{equation}
 {\cal A}'_i(l) = {\cal B}'_i(l) \; {\cal B}_i(l)^{-1} \; {\cal A}_i(l).
\end{equation}
${\cal B}'_i(l)$ denotes the matrix (\ref{Bmatrix}) with 
$\sigma_{ij}(l)$ replaced by $\sigma'_{ij}(l)$.
Note that it is not necessary to perform the whole product of $L$ matrices
but only a few $2\times 2$ matrix operations for each spin flip.
If accepted the new matrices ${\cal A}'_i(l)$ replace the old ones.
\item Move to the next time slice, $l+1$, by calculating
\begin{equation}
{\cal A}_i(l+1) =  {\cal B}_i(l)^{-1} {\cal A}_i(l) {\cal B}_i(l),
\end{equation}
for each lattice site $i$.
\item Move to (2).
\end{enumerate}
After $L$ cycles (2-4) one sweep through the $d+1$ dimensional system
is complete. It takes $\propto NL (Z/2)^2$ multiplications.
Step 3 leads to round-off errors, in particular at large $\beta$,
so one has to recompute the matrices ${\cal A}_i(l)$ from scratch from to time,
typically after ten time slices.

We note that the systematic error due to the Trotter decomposition
can be strongly reduced by a third order decoupling.
\begin{equation}
e^{\Delta\tau (A+B)} \approx 
e^{\Delta\tau A/2} \; e^{\Delta\tau B} \;e^{\Delta\tau A/2} + 
{\cal O} (\Delta\tau^3).
\end{equation}
While the leading correction in expectation values of hermitian operators
is ${\cal O} (\Delta\tau^2)$ for both second and third order decoupling,
the prefactors are typically a lot smaller in the latter case.
The implementation is simple. Formally the matrices (\ref{Bmatrix}) 
are changed to
\[
\tilde {\cal B}_i(l) \equiv  e^{\Delta\tau B_i S_i^x/2} \;
\left( \prod_{j\in {\rm NN}(i)} 
e^{\lambda_{ij}\sigma_{ij}(l) S_i^z}\right)
\; e^{\Delta\tau B_i S_i^x/2}.
\]

In the product (\ref{Amatrix}), however, there are always two
of such factors, $\exp({\Delta\tau B_i S_i^x/2})$, adding to 
$\exp({\Delta\tau B_i S_i})$, and the remaining factor at the beginning
of the product can be shifted to the end since the trace does not change under
cyclic permutation.
Hence, the Monte Carlo procedure remains completely unchanged, and
it is sufficient to replace each \em local operator \em ${\cal O}_i$ by
\begin{equation}
\tilde {\cal O}_i \equiv e^{-\Delta\tau B_i S_i^x/2} \;
{\cal O}_i \; e^{\Delta\tau B_i S_i^x/2}.
\end{equation}
The computational effort for these $2N$ additional $2 \times 2$
matrix multiplications is negligible.

\subsection{Observables}

The algorithm allows for the measurement of a variety
of static and time-dependent observables.
Since measurements for successive time slices are in general correlated
they are performed after every full sweep over space and time.

Interestingly, in the weak coupling limit ($J_{ij}=0$), all expectation 
values become exact, independent on the auxiliary field configuration,
i.e.~without any sampling.
This is not the case if one samples over the original Ising spins.
By analogy the auxiliary field approach for fermions\cite{DET}
exactly solves the noninteracting problem without sampling, while
world--line approaches\cite{WL} do not, and still require a full
Monte Carlo simulation to get observables.

\subsubsection{Static correlation functions}

Most static observables can be expressed in terms of 
local magnetizations and static correlation functions.
The components of the local magnetization $S_i^\alpha$ are given by
\begin{eqnarray}
\langle S_i^\alpha \rangle & = & 
\frac{1}{{\cal Z}}  \sum_{\{\sigma_{ij}(l)\}} \left( \prod_{k\neq i} 
{\rm Tr}  \, {\cal A}_k(1) \right) \; {\rm Tr} \,\{S_i^\alpha {\cal A}_i(1)\} 
\nonumber\\
& = &  \frac{1}{{\cal Z}}  \sum_{\{\sigma_{ij}(l)\}} w(\{\sigma_{ij}(l)\})\;
\frac{{\rm Tr} \,\{S_i^\alpha {\cal A}_i(1)\}}
{{\rm Tr} \, {\cal A}_i(1) }
\label{Mag}
\end{eqnarray}
with the weight function $w(\{\sigma_{ij}(l)\})$ from Eq.~(\ref{weight}).
That means we have to sum up the ratio of traces on the r.h.s.~of
(\ref{Mag}) over the auxiliary field configurations.
Similar equations hold for static correlation functions. In short:

\begin{eqnarray}
\langle S_i^\alpha \rangle & = & \left\langle
\frac{{\rm Tr} \,\{S_i^\alpha {\cal A}_i(1)\}} {{\rm Tr} \, {\cal A}_i(1)}
\right\rangle_{w} \\
\langle S_i^\alpha S_k^\beta \rangle & = & \left\langle
\frac{{\rm Tr} \,\{S_i^\alpha {\cal A}_i(1)\} \;
{\rm Tr} \,\{S_k^\beta {\cal A}_k(1)\}} 
{{\rm Tr} \, {\cal A}_i(1) \; {\rm Tr} \, {\cal A}_k(1)}
\right\rangle_{w}  {\scriptsize (i\neq k)} \\
\langle S_i^\alpha S_i^\beta \rangle & = & \left\langle
\frac{{\rm Tr} \,\{S_i^\alpha S_i^\beta  {\cal A}_i(1)\}}
 {{\rm Tr} \, {\cal A}_i(1)},
\right\rangle_{w} 
\end{eqnarray}
where $\langle\cdots\rangle_w$ stands for the sum over $\{\sigma_{ij}(l)\}$
configurations with proper weight.
  
\subsubsection{Static Susceptibilities}

Susceptibilities, in general, require the calculation
of correlation functions in imaginary time.
The homogeneous susceptibility for spin component $\alpha$ is defined as

\begin{equation}
\chi^\alpha=N \left [ 
\int_0^\beta \, d\tau \, \langle M^\alpha(\tau) M^\alpha(0)  
\rangle -\beta \langle M^\alpha \rangle^2 \right],
\label{chi}
\end{equation}
with the magnetization operator
\begin{equation}
M^\alpha(\tau)= e^{-\tau H} M^\alpha e^{\tau H},\;\;
 M^\alpha = \frac{1}{N} \sum_{i=1}^N S^\alpha_i.
\label{Mtau}
\end{equation}
In our discrete time approach the integral is replaced by a sum
over time slices, yielding:
\begin{eqnarray}
\chi^\alpha & =& \Delta\tau N \sum_{l=1}^L \left\langle  
M^\alpha(\Delta\tau l) \,
M^\alpha(0) \right\rangle - \beta N \langle M^\alpha \rangle^2\nonumber \\
& = & \frac{\Delta\tau}{N} \sum_{l=1}^L \sum_{n,m} \left\langle
S_m^\alpha(l) \, S_n^\alpha(0)  \right\rangle
 - \beta N \langle M^\alpha \rangle^2,
\end{eqnarray}
where $S_m^\alpha(l)\equiv S_m^\alpha(\Delta\tau l)$ means:

\begin{equation}
S_m^\alpha(l) =\left(\prod_{l'=1}^l B(l)'\right) \; S_m^\alpha \; 
\left(\prod_{l'=1}^l B(l)' \right)^{-1}.
\end{equation}

For the time-dependent correlation function we again employ the fact that
for a given auxiliary-field configuration operators for different
lattice sites commute and we obtain for $m\neq n$:

\begin{eqnarray}
\chi^\alpha_{mn} (l) & \equiv &
\langle S_m^\alpha(l) S_n^\alpha(0) \rangle \nonumber \\
& = & \left\langle e^{-(\beta- \Delta\tau l) H}\, S_m^\alpha 
\,e^{-\Delta\tau l H} \, 
 S_n^\alpha \, \right\rangle  \nonumber \\
& = & \frac{1}{{\cal Z}} \sum_{\{\sigma_{ij}\}} 
 \hspace{-.4em}\left( \hspace{-.2em}
\prod_{i\neq m,n} 
 {\rm Tr} \, {\cal A}_i(1)   \hspace{-.2em} \right) \hspace{-.2em}
{\rm Tr} \, \left[S_m^\alpha (l)\right] 
\;{\rm Tr} \, \left[S_n^\alpha (0)\right]   \nonumber \\
&=&  \left\langle
\frac{{\rm Tr} \, \left[S_m^\alpha (l)\right] \;
{\rm Tr} \, \left[S_n^\alpha (0)\right] } 
{{\rm Tr} \, {\cal A}_m(1) \; {\rm Tr} \, {\cal A}_n(1) }
\right\rangle_{w} \label{chitau},
\label{dynsusc}
\end{eqnarray}
where the operators ($2\times 2$ matrices) in brackets are defined as:

\begin{eqnarray}
\left[S_m^\alpha (l)\right] & = & S_m^\alpha (l)\, A_m(1) \nonumber \\
& = &\prod_{l'=1}^l {\cal B}_m(l')\; S_m^\alpha\;
\prod_{l'=l+1}^{L} {\cal B}_m(l') 
\label{Stau}
\end{eqnarray}
For the on-site correlation function we obtain similarly:
\begin{equation}
\chi^\alpha_{mm} (l) = \left\langle
\frac{{\rm Tr} \,\{ \left[S_m^\alpha (l)\right] \; S_m^\alpha\}} 
{{\rm Tr} \, {\cal A}_m(1)}
\right\rangle_{w}.
\end{equation}

Eq.~(\ref{Stau}) can be written as a product 
${\cal L}_m(l) S_m^\alpha {\cal R}_m(l)$
where the matrix products on the left and right hand side
of the spin operator are calculated iteratively:

\begin{eqnarray}
{\cal L}_m(1) \equiv  {\cal B}_m(1), \phantom{1m} & &
{\cal L}_m(l+1) =  {\cal L}_m(l) {\cal B}_m(l+1)  \\
{\cal R}_m(L) \equiv  1, \phantom{{\cal B}_m(1)1  } & &
{\cal R}_m(l-1) =  {\cal B}_m(l) {\cal R}_m(l) .
\end{eqnarray}

For each value of $l$, the product ${\cal L}_m(l) {\cal R}_m(l)=A_m(1)$.
To check for round-off errors this equality is tested from time to
time. No significant deviations were found for the parameters used.

$\chi^\alpha_{mn}$ also determines the dynamical susceptibility
in imaginary time from which, in principle, the real time dynamics 
can be extracted by an analytic continuation.

\begin{figure}
  \epsfxsize=7cm \vspace*{10mm}\hspace*{3mm} \epsfbox{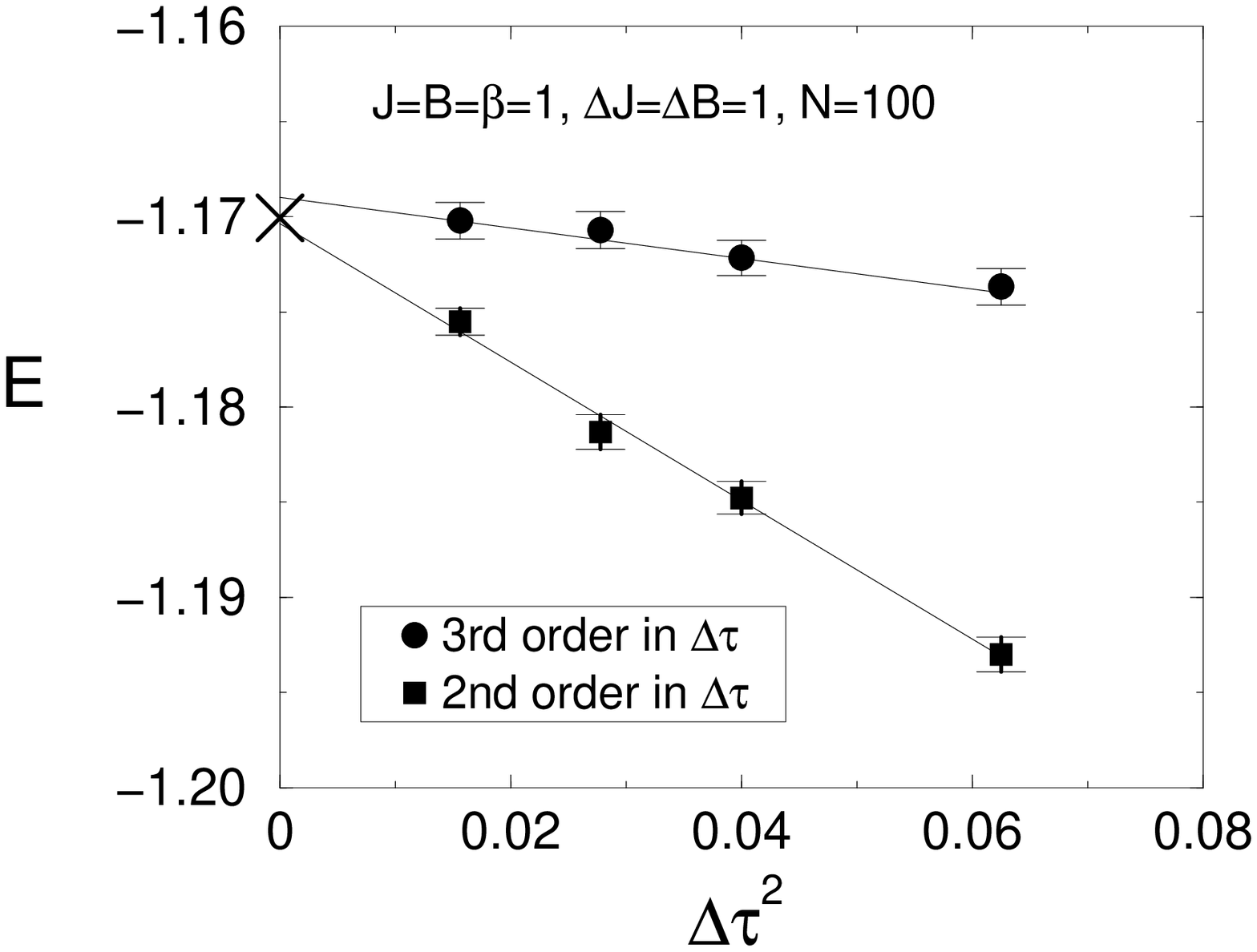}

  \epsfxsize=7cm \vspace*{-5mm}\hspace*{3mm} \epsfbox{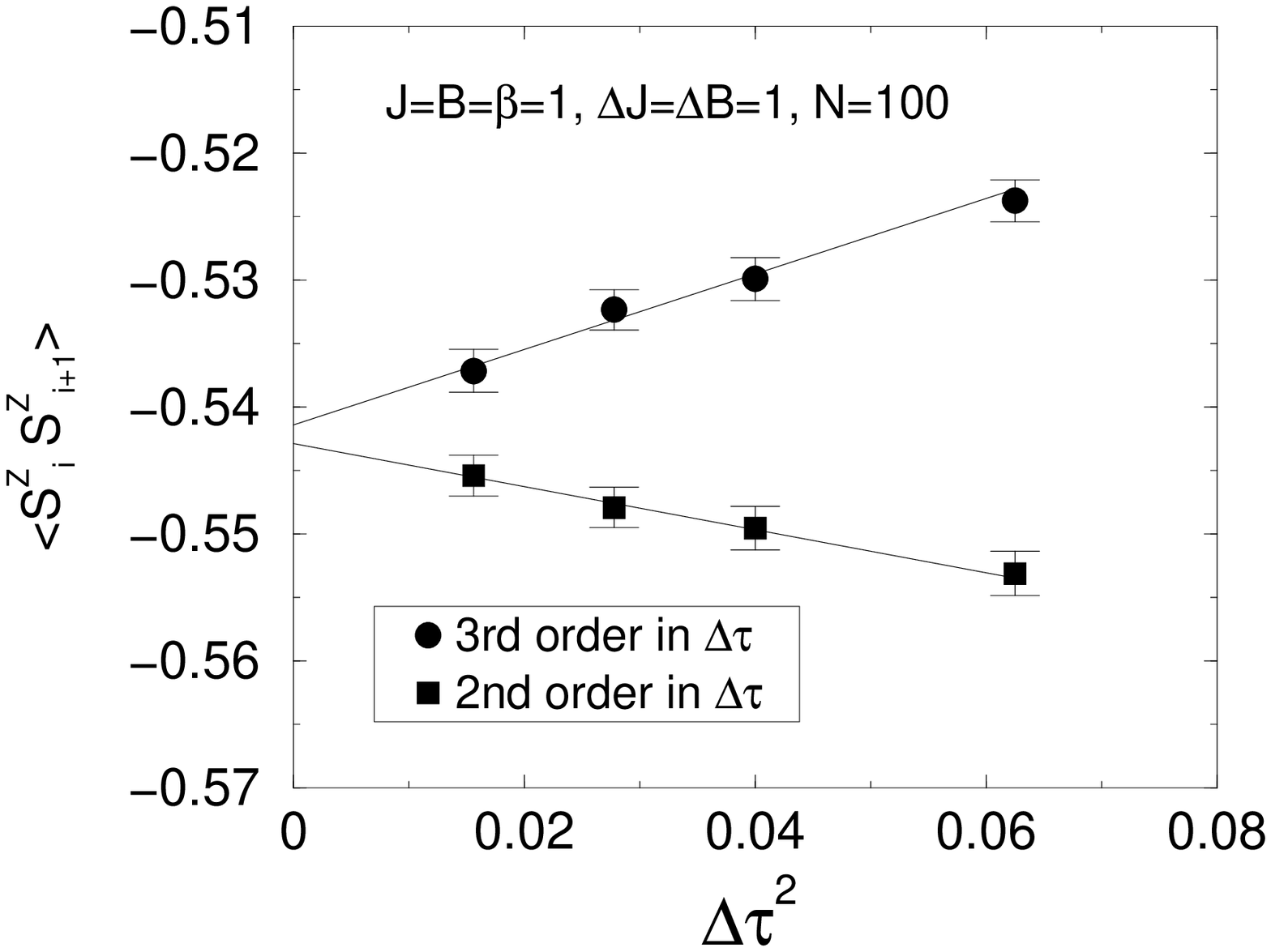}

  \epsfxsize=7cm \vspace*{-5mm}\hspace*{3mm} \epsfbox{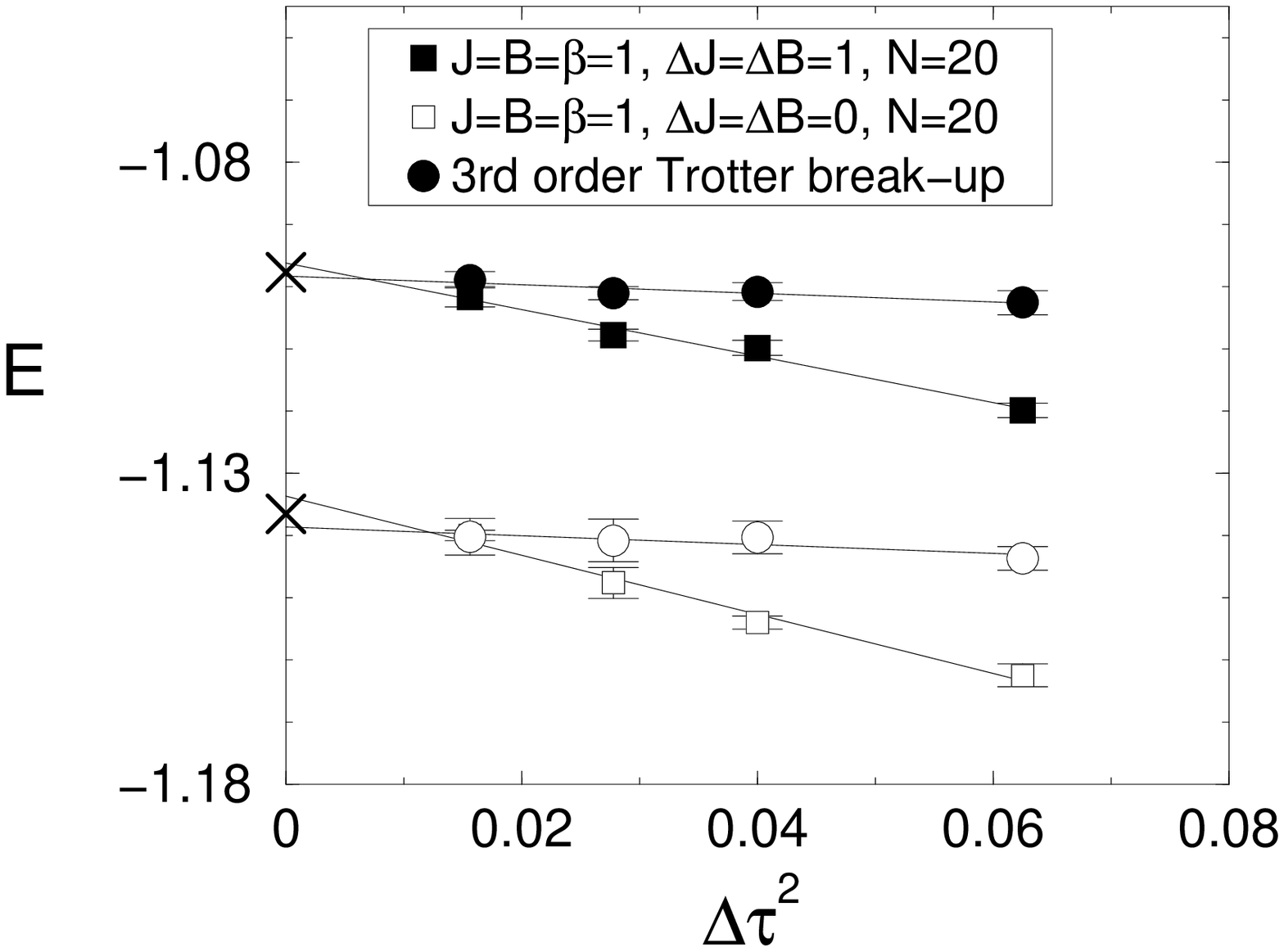}
\caption{Random transverse Ising chain with one disorder configuration.
2nd and 3rd order Trotter decomposition.
Exact results $(\times)$ from Jordan-Wigner transformation.}
\label{FigJW} 
\end{figure}

\subsection{Results}

\subsubsection{Random field, random bond transverse Ising chain} 

The one-dimensional Ising model in a transverse field
can be solved exactly using the
Jordan-Wigner transformation \cite{Lieb61,Pfeuty70}. 
With random bonds and/or random magnetic fields,
explicit formulas for finite open chains at finite temperatures were given
by Young \cite{Young97} which can be used to test the algorithm.
We did simulations of open chains of 20 and 100 sites
with one disorder configuration, and calculated
energy and nearest neighbor correlation functions (see Fig.~\ref{FigJW}).
The energy values are compared with the numerically exact ones.
The convergence with $\Delta\tau^2$ to the exact values is quite good.
The third order Trotter break-up leads to significantly smaller 
systematic errors in the energy. In the spin-spin correlation function,
however, the prefactor is somewhat larger.

\subsubsection{2D Transverse Ising Model}

\begin{figure}
  \epsfxsize=8cm \epsfbox{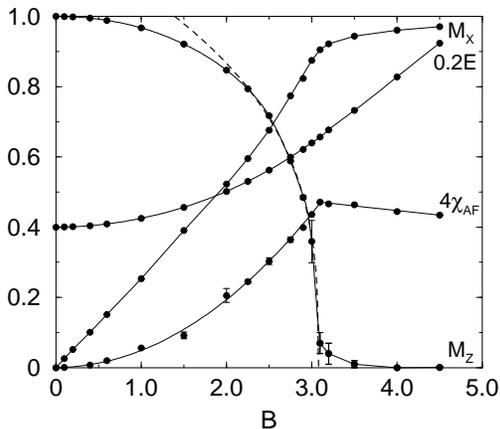}
\caption{
Longitudinal and transverse magnetization, $M_z$ and $M_x$, staggered 
susceptibility, $\chi_{AF}$, and energy $E$ vs.~transverse magnetic field $B$.
$10 \times 10$ lattice sites, $\beta=10$ with $\Delta\tau\to 0$ 
extrapolation.
Dashed line: fit $M_{\rm z}\propto (B_c-B)^\beta$ with $B_c(T=0)=3.08$.}
\label{Fig2Dpure}
\end{figure}

A second application of the algorithm is the phase transition
induced by a transverse magnetic field in the pure (non-random)
two-dimensional ferromagnetic Ising model. Fig.~\ref{Fig2Dpure}
shows results for one fixed system size at an inverse temperature $\beta=10$.
All quantities are extrapolated to $\Delta\tau\to 0$.

The phase transition is visible in the longitudinal and transverse
magnetization, $M_z$ and $M_x$, as well as in the staggered 
susceptibility, $\chi_{AF}$, which shows a kink at the transition.
The homogeneous susceptibility, however, was too strongly fluctuating
to give reliable results. The energy, $E$, behaves smoothly at the transition
as expected for a second order transition.
$M_z$ vanishes around $B\approx 3.1$.
If we assume that the finite system is essentially in its
ground state and take the value of the critical magnetic field
from high temperature expansions \cite{Elliot71,Oitmaa77}, $B_c=3.08$,
then we can fit $M_z$ near the transition by a power law and get
the exponent $\beta=0.32 \pm 0.01$. 
This is in remarkably good agreement with the value
for the classical $d+1=3$ dimensional Ising model, $\beta= 0.325 \pm 0.0015$
\cite{Zinn-Justin80} which should apply at the quantum critical point ($T=0$). 

Auto-correlation times are short even close to the transition point.
At $B=3.1$ we observe auto-correlations in $M_z$, $E$, and $\chi_{AF}$
below 0.1 after one sweep, and in $M_x$  and $\chi_{F}$ after four sweeps.
The data for Fig.~\ref{Fig2Dpure} were obtained within
about one day of computer time on a workstation.

\section{The Boson--Hubbard (XXZ) Model}

\subsection{Hamiltonian and Algorithm}

The boson Hubbard model is
\begin{eqnarray}
H &&= -t \sum_{\langle ij \rangle} (a_i^{\dagger} a_j^{\phantom\dagger}
+ a_j^{\dagger} a_i^{\phantom\dagger} ) \nonumber \\
    +V \sum_{\langle ij \rangle} n_i n_j 
&&+ U \sum_{i} n_i(n_i-1)
- \mu \sum_i n_i.
\label{bosonHM}
\end{eqnarray}
Here $a_i^{\phantom\dagger}$ and $a_i^\dagger$ are the 
destruction and creation operators for bosons on site $i$, 
and $n_i=a_i^\dagger a_i^{\phantom\dagger}$
is the number operator.  The first term is the kinetic energy, 
and $U$ and $V$ are on--site and
near--neighbor repulsions between bosons.
$\mu$ is a chemical potential which controls the 
density of electrons per site, $\langle n \rangle$, on the lattice.

To briefly illustrate some of the properties of
the model, consider the ground state phase diagram at $V=0$.  
As the chemical potential is raised,
the density of bosons on the lattice increases smoothly from zero.
However, if $U$ is large, then when the density goes through
$\langle n \rangle = 1$,
the chemical potential takes a sudden jump, since
at that point lattice sites become doubly occupied at the
cost of the big on--site repulsion $U$.
Similar jumps occur at all integer fillings.
These jumps represent a gap in the many--body energy spectrum, and
reflect the fact that the system is insulating at strong coupling.
If $U$ is sufficiently small, the kinetic energy dominates, and
the gap vanishes.  
Therefore, as $U/t$ is changed, the boson--Hubbard Hamiltonian 
undergoes a quantum phase transition between superfluid and insulating states.
Away from integer filling, the system is a superfluid at any ratio of $U/t$.
Nonzero $V$, or the introduction of disorder, likewise
allow for interesting new phases at $T=0$.

In the hard--core limit, the occupations are $n_i=0,1$.
Then $U$ drops out of the problem and
with the usual mappings
\begin{eqnarray}
&& a_i^\dagger \rightarrow S_i^+  = 
 \left(S_i^x + i S_i^y \right)/2 \nonumber \\
&& a_i^{\phantom\dagger} \rightarrow S_i^- =
\left(S_i^x + i S_i^y \right)/2 \nonumber\\
&& n_i \rightarrow S_i^z + 1/2,
\end{eqnarray}
Eq.~(29) transforms into the spin--1/2 XXZ model,
\begin{eqnarray}
H = && -t/2 \sum_{\langle ij \rangle} (S_i^x S_j^x + S_i^y S_j^y)  \nonumber\\
    && +V \sum_{\langle ij \rangle} (S_i^z +\frac12) 
(S_j^z +\frac12)   \nonumber \\
&& - \mu \sum_i (S_i^z+\frac12)
\end{eqnarray}
An occupied site is hence identified with a spin up, etc., and the
issues discussed above for the boson--Hubbard model can be reformulated
in spin language.  For example, the competition between superfluid and
insulating phases corresponds to the between magnetic order in the
XY and Z directions.

As mentioned above, the 
auxiliary-field
decoupling (\ref{HStrafo}) can be performed
for different spin components independently.
In the case of the $y$ component it is useful to extract a trivial 
factor of $i$ in order to avoid complex matrix elements.
Alternatively, by using the relations
\begin{eqnarray} \label{asquare}
(a^\dagger)^2 = a^2 & = &  0  \nonumber\\
 a^\dagger a + a a^\dagger & = & 1  ,
\end{eqnarray}
valid in the hard-core limit, one can 
directly decouple the kinetic energy term in (\ref{bosonHM}):

\begin{eqnarray}
e^{\Delta\tau t a^\dagger_i a^{\phantom \dagger}_j} 
& =& 1+\Delta\tau \, t \, a^\dagger_i a^{\phantom \dagger}_j 
\nonumber \\
& = & 1+ \frac{\Delta\tau}{2} \, t \,
(a^\dagger_i+ a^{\phantom \dagger}_j )^2  \nonumber \\
& = & \cosh \left[\sqrt { \Delta\tau \, t } \, 
(a^\dagger_i+ a^{\phantom \dagger}_j )\right]\nonumber \\
& = & \frac{1}{2} \sum_{\sigma=\pm 1} 
e^{\sqrt { \Delta\tau \, t }\,(a^\dagger_i+ a^{\phantom \dagger}_j ) \sigma}.
\label{HStrafo2}
\end{eqnarray}
The same decoupling is used for the hermitian conjugated term
$a^\dagger_j a^{\phantom \dagger}_i $ yielding, together with
the $S_i^z S_j^z$ decoupling, three auxiliary Ising-type fields
for the XXZ model 
which can be treated in analogy to the algorithm described in Sec.~II. 

\subsection{Results}

\begin{figure}
  \epsfxsize=8cm \epsfbox{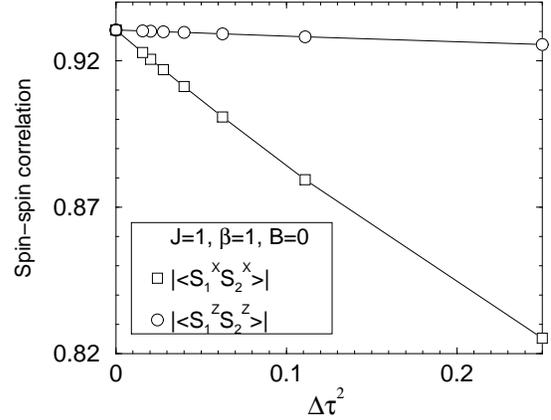}
\caption{AF Heisenberg model on two sites.
Full enumeration over all possible auxiliary field configurations.
$\Delta\tau$ extrapolation with 2nd order Trotter decomposition.}
\label{FigXY}
\end{figure}

In order to check the algorithm and the code we calculate the 
nearest-neighbor spin-spin correlation function
of the isotropic antiferromagnetic (AF) Heisenberg model
on a $2\times 2$ lattice. First, we did a full enumeration 
over all possible auxiliary spin configuration for several
values of $\Delta\tau$. Fig.~\ref{FigXY} shows the 
transverse and longitudinal correlation function vs.~$\Delta\tau^2$.
Apparently they converge quadratically with $\Delta\tau$ to the exact
value, however with different prefactors.

\begin{figure}
  \vspace{-5mm}
  \epsfxsize=8cm \epsfbox{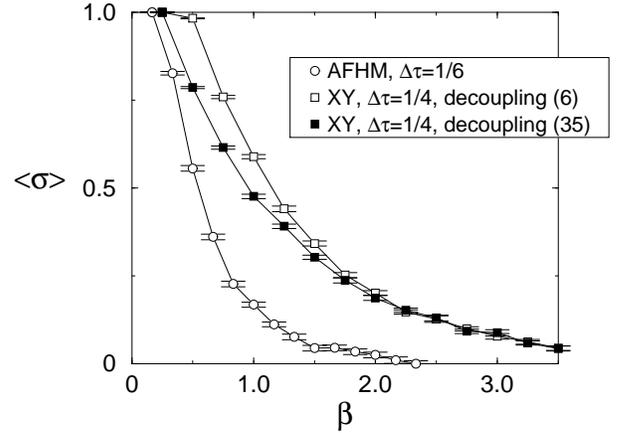}
\caption{Average sign of the weight function for 
AF Heisenberg and XY models on four lattice sites
$(J=1, B=0)$. 
$\sigma$ decreases exponentially with inverse temperature $\beta$.}
\label{Sign}
\end{figure}

Even for very small systems at relatively high temperature
severe minus-sign problems occur. 
Fig.~\ref{Sign} shows the average sign $\langle\sigma\rangle$ 
of $w$ vs.~$\beta$ for systems with four sites. 
$\langle\sigma\rangle$ vanishes exponentially with $\beta$.
Values of $\langle\sigma\rangle$
below approx.~0.2 preclude an efficient Monte Carlo sampling.
The values of $\langle\sigma\rangle$ do not much differ
for the two different decouplings, (\ref{HStrafo}) and (\ref{HStrafo2}), 
of the XY contribution. 

Hence the algorithm does not appear to be suitable
for models with couplings in more than one spin component.

\section{CONCLUSIONS}

We have formulated an auxiliary field Quantum Monte Carlo
algorithm for spin and hard--core boson systems.
In boson language, it is based on a Hubbard--Stratonovich
decoupling of the kinetic energy term, leaving a set of independent
one--site problems in a fluctuating external field.
Such a procedure has been used in analytic studies of the
boson--Hubbard model \cite{FISHER}.
The algorithm scales linearly with the spatial lattice size, and inverse
temperature, sharing that attractive feature of world--line
approaches compared to fermion auxiliary field techniques. 
However, unlike traditional world--line techniques it has very
short auto--correlation times.

Unfortunately, like fermion auxiliary field approaches, the determinants
can go negative, resulting in a sign problem.  We showed in the
case of the Ising model in a transverse field that an appropriate
spin rotation can eliminate the problem, making our approach a valuable
one for studying this problem in more that (1+1) dimensions,
where the Jordan--Wigner approach does not work.
Traditionally formulated world--line simulations,
which would map the problem onto a highly anisotropic
classical Ising model, suffer from large auto--correlation
times that are absent in the present algorithm.
Finally, our approach shows promise for the extraction of
dynamical correlation functions, a key problem in understanding
glassy dynamics in the random field case.


\begin{references}

\bibitem{LOOP}
H.~G.~Evertz, in \em Numerical Methods for Lattice Quantum Many-Body
Problems, \em ed. D.~J.~Scalapino, Addison-Wesley-Longman (1998),
and references cited therein.

\bibitem{CONTTIME}
B.B.~Beard and U.J.~Wiese, Phys.~Rev.~Lett.~{\bf 77}, 5130 (1996);
and N.V.~Prokofev, B.V.~Svistunov, and I.S.~Tupitsyn,
J.~Expt.~and Theor.~Phys.~{\bf 87}, 310 (1998).

\bibitem{DET}
R.~Blankenbecler, D.J.~Scalapino, and R.L.~Sugar,
Phys.~Rev.~D {\bf 24}, 2278 (1981).

\bibitem{YOUNG}
D.~S.~Fisher, Phys.~Rev.~B {\bf 51} 6411 (1995);
A.~P.~Young, Phys.~Rev.~B {\bf 56} 11691 (1997); and
D.~S.~Fisher and A.P.~Young, 
Phys.~Rev.~B {\bf 58}, 9131 (1998).

\bibitem{FISHER}
M.~P.~A.~Fisher, P.~B.~Weichman, G.~Grinstein, and D.~S.~Fisher,
Phys.~Rev.~B {\bf 40}, 546 (1989).
 
\bibitem{Chakrabarti96}
For a review see: B.~K.~Chakrabarti, A.~Dutta, and P.~Sen,
\em Quantum Ising Phases and Transitions in
Transverse Ising Models; \em Springer (Berlin 1996).

\bibitem{HIRSCH} J.~E.~Hirsch, Phys.~Rev.~B {\bf 31}, 4403 (1985).

\bibitem{WL} J.~E.~Hirsch, R.~L.~Sugar, D.~J.~Scalapino and R.~Blankenbecler,
Phys.~Rev.~B {\bf 26}, 5033 (1982).

\bibitem{Lieb61}
E.~Lieb, T.~Schulz and D.~Mattis, Ann.~Phys.~(NY), {\bf 16}, 407 (1961).

\bibitem{Pfeuty70}
P.~Pfeuty, Ann.~Phys.~(NY), {\bf 27}, 79 (1970).

\bibitem{Young97}
A.~P.~Young, Phys.~Rev.~B {\bf 56}, 11691 (1997).

\bibitem{Elliot71} 
R.~Elliot and C.~Wood, J.~Phys.~C {\bf 4}, 2359 (1971).

\bibitem{Oitmaa77}
J.~Oitmaa and M.~Plischke, Physica B {\bf 86-88}, 577 (1977).

\bibitem{Zinn-Justin80} 
J.~C.~Le Guillou and J.~Zinn-Justin, Phys.~Rev.~B {\bf 21}, 3976 (1980).

\end{references}
\end{document}